\documentclass[runningheads]{llncs}
\usepackage[T1]{fontenc}
\usepackage{graphicx}
\usepackage[hidelinks]{hyperref}
\usepackage{subcaption}
\usepackage{multirow}

\usepackage{amsmath}
\usepackage{amssymb}
\usepackage{makecell}
\usepackage{diagbox}
\usepackage{pifont}
\newcommand{\cmark}{\ding{51}}
\newcommand{\xmark}{\ding{55}}
\usepackage{balance}
\usepackage{ifthen}
\usepackage[table]{xcolor}
\usepackage{booktabs}
\usepackage{cite}

\usepackage{scalerel}
\usepackage{tikz}

\newcommand*\emptycirc[1][1ex]{\tikz\draw (0,0) circle (#1);} 
\newcommand*\halfcirc[1][1ex]{%
  \begin{tikzpicture}
  \draw[fill] (0,0)-- (90:#1) arc (90:270:#1) -- cycle ;
  \draw (0,0) circle (#1);
  \end{tikzpicture}}
\newcommand*\fullcirc[1][1ex]{\tikz\fill (0,0) circle (#1);}

\usepackage[strict]{changepage}
\usepackage{framed}

\newboolean{showcomments}
\setboolean{showcomments}{true}
\ifthenelse{\boolean{showcomments}}
 { \newcommand{\mynote}[2]{
      \fbox{\bfseries\sffamily\scriptsize#1}
        {\small$\blacktriangleright$\textsf{\emph{#2}}$\blacktriangleleft$}}}
        { \newcommand{\mynote}[2]{}}

\begin{document}

\title{Inference Attacks on Encrypted Online Voting\\via Traffic Analysis}

\author{Anastasiia~Belousova\inst{1}\orcidID{0009-0005-7236-7527} \and Francesco~Marchiori\inst{1}\orcidID{0000-0001-5282-0965} \and 
Mauro~Conti\inst{1,2}\orcidID{0000-0002-3612-1934}}

\authorrunning{A. Belousova et al.}

\institute{University of Padova, Padova, Italy \and Örebro University, Örebro, Sweden\\
\email{anastasiia.belousova@studenti.unipd.it, francesco.marchiori@math.unipd.it, mauro.conti@unipd.it}}

\maketitle

\begin{abstract}
Online voting enables individuals to participate in elections remotely, offering greater efficiency and accessibility in both governmental and organizational settings. As this method gains popularity, ensuring the security of online voting systems becomes increasingly vital, as the systems supporting it must satisfy a demanding set of security requirements. Most research in this area emphasizes the design and verification of cryptographic protocols to protect voter integrity and system confidentiality. However, other vectors, such as network traffic analysis, remain relatively understudied, even though they may pose significant threats to voter privacy and the overall trustworthiness of the system.

In this paper, we examine how adversaries can exploit metadata from encrypted network traffic to uncover sensitive information during online voting. Our analysis reveals that, even without accessing the encrypted content, it is possible to infer critical voter actions, such as whether a person votes, the exact moment a ballot is submitted, and whether the ballot is valid or spoiled. We test these attacks with both rule-based techniques and machine learning methods. We evaluate our attacks on two widely used online voting platforms, one proprietary and one partially open source, achieving classification accuracy as high as 99.5\%. These results expose a significant privacy vulnerability that threatens key properties of secure elections, including voter secrecy and protection against coercion or vote-buying. We explore mitigations to our attacks, demonstrating that countermeasures such as payload padding and timestamp equalization can substantially limit their effectiveness.

\end{abstract}

\section{Introduction}

Protecting ballot secrecy is a cornerstone of democratic elections, ensuring voters can express their preferences freely without fear of coercion, retaliation, or undue influence.
Many countries enforce this through strict legal measures designed to prevent any traceability of the vote.
For instance, photographing marked ballots is prohibited in Germany~\cite{ph-germany}, Ireland~\cite{ph-ireland}, Brazil~\cite{ph-brazil}, and numerous U.S. states~\cite{ph-usa}, as such images could compromise the confidentiality of the ballot.
These legal safeguards underscore a shared principle: the act of voting must remain private and unverifiable by others, preserving both voter anonymity and the integrity of the electoral process.
The risks of violating these guarantees are not hypothetical.
In Spain’s 2023 local elections, police uncovered networks exchanging postal votes for cash and favors~\cite{elect-spain}.
Similar concerns have emerged in Georgia and Bulgaria, where coercion has targeted public-sector workers and marginalized communities~\cite{elect-georgia, elect-bulgaria}.
In rare but alarming cases, such as Russia’s 2024 presidential election, voters who submitted spoiled ballots with anti-war messages were fined or arrested after their ballots were visible through transparent boxes~\cite{spb-case, tula-case}.
These examples show how even subtle violations of secrecy can lead to real-world harm.

As elections move online, these challenges take on new forms.
Online voting has gained traction in both governmental and organizational settings.
Countries like Estonia and Switzerland have institutionalized digital voting, while private entities increasingly adopt online platforms for shareholder meetings and union decisions.
This shift is driven by convenience, remote participation, and faster vote processing—advantages that became particularly clear during the COVID-19 pandemic~\cite{licht2021vote}.
Technologically mediated elections are described using overlapping terms: \textit{electronic}, \textit{digital}, and \textit{online} voting.
Electronic voting refers broadly to the use of devices at polling stations.
Digital voting builds on this with end-to-end software infrastructure.
Online voting, in turn, involves casting ballots remotely via the internet, typically through secure platforms or apps~\cite{format-diff}.
While offering clear benefits in accessibility and turnout, online voting must also meet the high standards of secrecy and trust long established by traditional paper-based methods.

With this transition comes the expectation that online voting systems must meet and exceed the integrity and trustworthiness of traditional paper-based elections.
Paper voting, despite its physical limitations, has long been regarded as the gold standard for ensuring transparency and public confidence.
Voters cast their ballots in private booths, often under the supervision of independent observers, and the process is verifiable at every stage.
If online voting is to be a viable alternative, it must preserve not only convenience but also uphold fundamental democratic guarantees, particularly voter privacy and protection from manipulation.
Recent advancements in online voting have predominantly focused on enhancing security through sophisticated cryptographic protocols and the integration of blockchain technologies~\cite{jafar2021blockchain}.
These efforts aim to ensure vote integrity, authentication, and transparency, leveraging mechanisms such as homomorphic encryption~\cite{sheela2021voting}, zero-knowledge proofs~\cite{wu2023smart}, and decentralized ledgers~\cite{lalitha2022decentralized}.
While these developments address critical aspects of secure voting, they often overlook the vulnerabilities inherent in the network infrastructure itself.
Specifically, the potential for adversaries to exploit metadata to infer sensitive information about voter behavior remains underexplored.

\paragraph{Contributions.}
This paper presents the first in-depth analysis of metadata-based inference attacks on two real-world online voting platforms.
We demonstrate that even without decrypting traffic content, a passive network adversary can extract sensitive information purely from encrypted metadata.
Our attacks reveal that adversaries with varying levels of access can determine whether a user voted, when the vote was cast, and whether the ballot was accepted or rejected.
These findings pose a serious threat to voter privacy.
The act of voting (or choosing not to vote) is itself a political signal, and the exposure of such actions can enable coercion or retaliation.
Similarly, although accidental spoiling is rare in online systems, deliberately spoiled ballots are often used as a form of protest and are publicly reported in many jurisdictions as indicators of political discontent.
The ability to infer such actions undermines key democratic principles, especially in contexts where voter pressure is a known issue, such as in Georgia and Bulgaria~\cite{elect-georgia, elect-bulgaria}.
Our contributions can be summarized as follows.\footnote{We conducted responsible disclosure with the companies involved. Samples of the code used in this paper are available upon request.}
\begin{itemize}
    \item We show that encrypted traffic alone can reveal, at a minimum, the moment of ballot submission, and at best, the full sequence of voter actions.
    \item We demonstrate that ballot validity status can be predicted using learning-based models trained on payload sizes and timing data.
    \item We evaluate our methodologies on two real-world online voting platforms: Eligo, a proprietary system, and POLYAS, a partially open-source solution. Our analysis achieves an accuracy of up to 99.5\%, demonstrating the high effectiveness and generalizability attacks.
    \item We propose and evaluate two countermeasures, quantifying their effectiveness as well as the trade-offs they introduce in terms of delay and memory overhead.
\end{itemize}

\paragraph{Organization.}
The remainder of this paper is structured as follows.
In Section~\ref{sec:related}, we review related work on online voting systems, focusing on their security requirements and recent developments.
Section~\ref{system-threat-model} introduces our system and threat model, outlining how adversaries can launch attacks without decrypting the payload.
We describe our attack methodology in Section~\ref{sec:methodology} and present its empirical evaluation in Section~\ref{sec:evaluation}.
Section~\ref{sec:discussion} discusses the implications of our findings and Section~\ref{sec:countermeasure} introduces our proposed countermeasures.
Finally, Section~\ref{sec:conclusions} concludes the paper.
\section{Related Works}
\label{sec:related}

We begin by reviewing prior work on online voting systems.
Section~\ref{subsec:21} outlines the fundamental security requirements for such systems and compares prominent publicly available and commercial platforms.
In Section~\ref{subsec:22}, we examine recent research directions and advances in the security of electronic voting.

\subsection{Online Voting Systems Security}
\label{subsec:21}

While the literature provides numerous evaluations of cryptographic protocols or specific classes of e-voting systems, such as blockchain-based platforms~\cite{blockchain-comp} or systems used for high-stakes national elections~\cite{gov-comp}, comparative analyses across diverse real-world deployments remain sparse.
Prior work often centers on algorithmic innovation or theoretical properties, overlooking how practical systems, used in binding elections, fulfill key security guarantees in practice.

\paragraph{Online Voting Systems.}
To help ground our study in realistic deployments, we include a focused comparative review of a small but diverse set of online voting platforms.
These systems differ in geographical scope, electoral application (ranging from organizational to governmental), and transparency models.
Although not the central aim of our work, this comparative lens offers useful context on how different platforms address core security requirements, based on publicly accessible information.
The platforms we examine include both open-source and proprietary systems, offering a diverse snapshot of real-world online voting solutions.
The following systems are considered:
Assembly Voting~\cite{website-AV}, Belenios~\cite{website-Belenios}, Civitas~\cite{civitas}, ElectionBuddy~\cite{website-EB}, Eligo~\cite{eligo-website}, Helios~\cite{website-Helios}, POLYAS~\cite{website-Polyas}, and Voatz~\cite{website-Voatz}.

\paragraph{Security Requirements.}
In our analysis, we focus on five widely recognized requirements~\cite{req-def, civitas}: 
\begin{itemize}
    \item \textit{Eligibility:} only eligible voters should be able to vote.
    \item \textit{Ballot secrecy:} no actor involved in the voting process should be able to link a ballot to a voter.
    \item \textit{Individual verifiability:} each voter can check that their own vote is included in the tally.
    \item \textit{Universal verifiability:} anyone can check that all votes cast are counted, that only authorized votes are counted, and that no votes are changed during counting.
    \item \textit{Coercion resistance:} voters cannot prove whether or how they voted, even if they can interact with the adversary while voting.
\end{itemize} 
The results of our comparative analysis are summarized in Table~\ref{tab:compar-analysis}.
Some security requirements, such as eligibility, anonymity, and privacy, are typically assumed to be fulfilled a priori and were explicitly claimed by all reviewed systems.
Others, like ballot secrecy, were uniformly stated as upheld.
However, this assumption-based trust model creates a blind spot: if a foundational property such as privacy is compromised, for instance, through side-channel or metadata-based inference attacks, then dependent guarantees like ballot secrecy and receipt-freeness may no longer hold in practice.
This fragility undermines higher-level protections such as coercion resistance and opens the door to vote-buying scenarios, even if the systems do not explicitly claim to defend against them.
Our work demonstrates how inference from encrypted traffic can trigger such cascading failures, raising concerns about the real-world robustness and trustworthiness of these systems.

\begin{table*}[!htbp]
\scriptsize
\renewcommand{\arraystretch}{1.2}
\centering
\caption{Comparison of selected online voting systems.}
\label{tab:compar-analysis}
\resizebox{\linewidth}{!}{%
\begin{tabular}{c|c|c|c|c|c|c|c|c}
\hline

\multirow{3}{*}{\textbf{System}} &
\multirow{3}{*}{\textbf{\begin{tabular}[c]{@{}c@{}}Cryptographic\\Algorithm\end{tabular}}} &
\multirow{3}{*}{\textbf{Certifications}} &
\multirow{3}{*}{\textbf{\begin{tabular}[c]{@{}c@{}}Open\\Source\end{tabular}}} & 
\multicolumn{5}{c}{\textbf{Security Requirements}} \\ \cline{5-9} 

& & & &
\textbf{Eligibility} &
\textbf{\begin{tabular}[c]{@{}c@{}}Ballot\\Secrecy\end{tabular}} &
\textbf{\begin{tabular}[c]{@{}c@{}}Individual\\Verifiability\end{tabular}} & 
\textbf{\begin{tabular}[c]{@{}c@{}}Universal\\Verifiability\end{tabular}} & 
\textbf{\begin{tabular}[c]{@{}c@{}}Coercion\\Resistance\end{tabular}} \\ \hline

\rowcolor{gray!15}
\begin{tabular}[c]{@{}c@{}}Assembly \\ Voting\end{tabular} & \begin{tabular}[c]{@{}c@{}}ECC, ElGamal,\\AES-GCM\end{tabular} & \begin{tabular}[c]{@{}c@{}}ISO 27001,\\ISAE 3000\end{tabular}  & Partially & \fullcirc & \fullcirc & \fullcirc  & \fullcirc &   \emptycirc \\ \hline

Belenios & ElGamal & N/A & \cmark & \fullcirc & \fullcirc & \fullcirc  & \fullcirc &    \emptycirc \\ \hline

\rowcolor{gray!15}
Civitas & RSA, ElGamal & N/A & \cmark & \fullcirc & \fullcirc & \fullcirc  & \fullcirc &   \fullcirc  \\ \hline

\begin{tabular}[c]{@{}c@{}}Election\\Buddy\end{tabular} & Not specified & Not specified  & \xmark &  \fullcirc & \fullcirc & \emptycirc  & \emptycirc &   \emptycirc \\ \hline

\rowcolor{gray!15}
Eligo & AES & \begin{tabular}[c]{@{}c@{}}ISO 27001\\ISO 9001:2015\end{tabular} & \xmark & \fullcirc  & \fullcirc & \fullcirc  & \fullcirc &   \halfcirc \\ \hline

Helios & ElGamal & N/A & \cmark & \fullcirc & \fullcirc & \fullcirc  & \fullcirc &    \emptycirc \\ \hline

\rowcolor{gray!15}
POLYAS & \begin{tabular}[c]{@{}c@{}}AES, RSA\\ECIES, ElGamal\end{tabular}  & \begin{tabular}[c]{@{}c@{}}ISO 27001, \\ BSI\end{tabular}  & Partially &  \fullcirc & \fullcirc  &  \fullcirc & \fullcirc &   \emptycirc   \\ \hline

Voatz & AES-GCM & Not specified & \xmark & \fullcirc & \fullcirc &  \emptycirc & \emptycirc &   \emptycirc  \\ \hline

\hline
\multicolumn{9}{l}{\footnotesize{\fullcirc: fulfilled, \halfcirc: partially fulfilled, \emptycirc: not fulfilled or not mentioned.}}
\end{tabular}
}
\end{table*}

To demonstrate and evaluate our proposed inference methodology, we focus on Eligo, the most feature-rich and prominent among the analyzed systems, and subsequently validate our approach on POLYAS, which offers similar functionality and voting flow but represents an independent platform. This selection strategy allows us to ground the attack in practical deployments and demonstrate its generalizability.

\subsection{Advances and Attacks Overview}
\label{subsec:22}

Recent advancements in cryptographic protocols have significantly enhanced the security and verifiability of electronic voting systems.
End-to-end verifiable (E2E) systems like Helios and Belenios enable voters to confirm that their votes are accurately recorded and tallied without compromising ballot secrecy.
Innovations such as D-DEMOS and Hyperion have introduced distributed architectures and coercion mitigation strategies, further strengthening the integrity of online voting platforms~\cite{chondros2016d, damodaran2024hyperion}.
Additionally, the integration of blockchain technology, as seen in systems like Voatz, aims to provide transparent and tamper-evident records of the voting process.
However, a comprehensive security analysis of Voatz by Specter et al. revealed significant vulnerabilities, including the potential for passive network adversaries to recover users' secret ballots, thereby undermining the system's integrity and privacy claims~\cite{specter2020ballot}.
This study underscores the necessity for transparency and rigorous security evaluations in the deployment of online voting platforms.
Further research has identified multiple privacy and integrity attacks on online voting systems.
Indeed, in other cases, inference and side‐channel attacks have also been demonstrated even when ballots are cryptographically protected.
For example, Brunet et al. show that subtle differences in confirmation-page content (e.g., packet lengths) can leak voters’ choices~\cite{brunet2022review}.
Blanchard et al. analyzed a real voting system (France’s Neovote) and found that its verification procedure could be subverted to completely breach the voters’ privacy in multiple configurations, effectively deanonymizing ballots~\cite{blanchard2022analysis}.
These studies illustrate that adversaries can exploit metadata (timing, packet sizes, or protocol behaviors) to infer sensitive information beyond the ballot contents.
To date, however, few works have systematically quantified metadata-based inference, making our analysis of such attacks a novel contribution.
\section{System and Threat Model}
\label{system-threat-model}

\begin{figure}[!htbp]
    \centering
    \includegraphics[width=.75\linewidth]{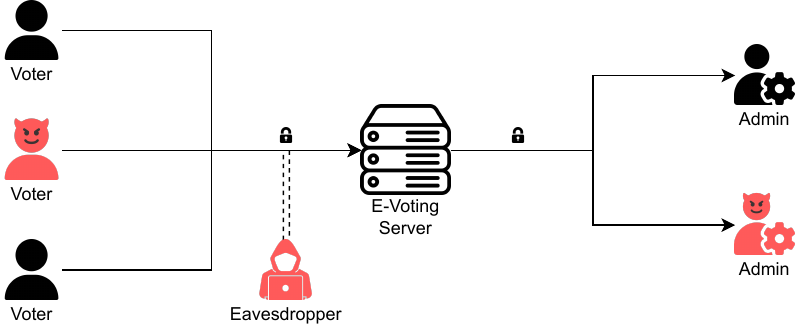}
    \caption{System and threat model overview.}
    \label{fig:model}
\end{figure}

We describe the operational context of a typical online voting system (Section~\ref{subsec:system}) and the capabilities of attackers attempting to infer sensitive information through encrypted traffic analysis (Section~\ref{subsec:threat}).
An overview of our considered system and threat model is shown in Fig.~\ref{fig:model}.

\subsection{System Model}
\label{subsec:system}

We examine a real-world online voting platform used in organizational settings such as university elections, academic societies, or corporate boards.
Our work builds on a well-established view of privacy in e-voting that includes metadata confidentiality, covering participation, timing, and ballot-type privacy.
These aspects are recognized in the literature~\cite{bernhard2017security} and official guidelines~\cite{coe2017evoting}.
The systems enable remote participation via web browsers or dedicated apps, with voters authenticating through credentials issued by their organizations.
Communication occurs over TLS, and platforms commonly implement eligibility checks, session management, and confirmation steps.
Election administrators oversee the process and may access metadata like session logs or timestamps, depending on the platform’s design.
This model reflects widely adopted systems such as Eligo and POLYAS, which ensure ballot secrecy and integrity through cryptographic mechanisms.
While the system is assumed uncompromised, side-channel leakage may silently undermine its security in realistic deployments where such vectors are not actively mitigated.

\subsection{Threat Model}
\label{subsec:threat}

We assume adversaries do not compromise cryptographic primitives, user credentials, or backend infrastructure. Instead, their objective is to infer sensitive information, such as whether and when a user voted, or whether their ballot was valid or spoiled, through traffic analysis and metadata observation. We consider three attacker roles with varying capabilities:
\begin{enumerate}
    \item \textbf{Voter:} A malicious voter observes their own client-side traffic during different stages (e.g., login, voting) and learns to recognize traffic patterns. They may then apply this knowledge to identify similar behaviors by others on the same network, inferring voting times or ballot status even without direct access.
    \item \textbf{Administrator:} A dishonest administrator has access to backend metadata (e.g.,~logs, timestamps) but not vote contents. Platforms like Eligo and POLYAS may expose real-time activity or detailed turnout data, which, when correlated with network traces, can reveal who cast valid or spoiled ballots.
    \item \textbf{Eavesdropper:} A passive network observer, such as an ISP, campus IT, or compromised router, can monitor TLS-encrypted traffic. While unable to see vote contents or user identities directly, they may infer behavior via traffic flow patterns and timing. In small networks, auxiliary data (e.g., DHCP logs) can further deanonymize users.
\end{enumerate}
These side-channel attacks do not violate cryptographic guarantees but exploit observable metadata. They are especially concerning in scenarios demanding strong anonymity or coercion resistance, where leaking vote timing or status can severely weaken electoral integrity.
\section{Methodology}
\label{sec:methodology}

This section presents the design of our framework, developed to identify and evaluate patterns in encrypted network traffic generated during electronic voting.
The framework is specifically tailored to classify voter actions and detect differences between valid and spoiled ballots based solely on metadata features observable from outgoing TLS packets.
Our approach incorporates pre-processing and feature engineering (Section~\ref{subsec:preprocessing}), voter action and submission identification (Sections~\ref{subsec:action} and~\ref{subsec:submission}), and classification of valid vs spoiled ballots (Section~\ref{subsec:valid}).

\paragraph{Tested Platforms.}
We focus our analysis on two widely used online voting platforms: Eligo and POLYAS.
Eligo has supported over 4,000 organizations~\cite{eligo-website}, while POLYAS has handled elections with more than 2 million voters annually~\cite{polyas-casestudies}.
It is also worth noting that Eligo was prioritized over Civitas (which satisfies more security requirements, as shown in Table~\ref{tab:compar-analysis}) due to its production use; Civitas, while valuable, remains largely a research prototype.
POLYAS is then added to validate generalizability across different real-world systems with similar workflows.
Both platforms are widely used in governmental and organizational elections and support the casting of spoiled or invalid ballots, a critical feature for evaluating whether traffic analysis can distinguish between voting outcomes.
The platforms differ in handling multi-ballot elections: Eligo allows selective participation per ballot, while POLYAS requires a decision on all ballots before submission.
To avoid inconsistencies, we limit our analysis to single-ballot elections, eliminating variability due to optional interactions.
Our study relies solely on encrypted TLS metadata.
From this, we extract timestamps and application data payload sizes to compute both raw and derived features, such as normalized timing (relative to the first packet of an action) and payload-based patterns.
These form the foundation for action segmentation and classification.

\subsection{Pre-Processing and Feature Engineering}
\label{subsec:preprocessing}

\begin{figure}[!htbp]
    \centering
    \begin{subfigure}[t]{0.24\linewidth}
        \includegraphics[width=\linewidth]{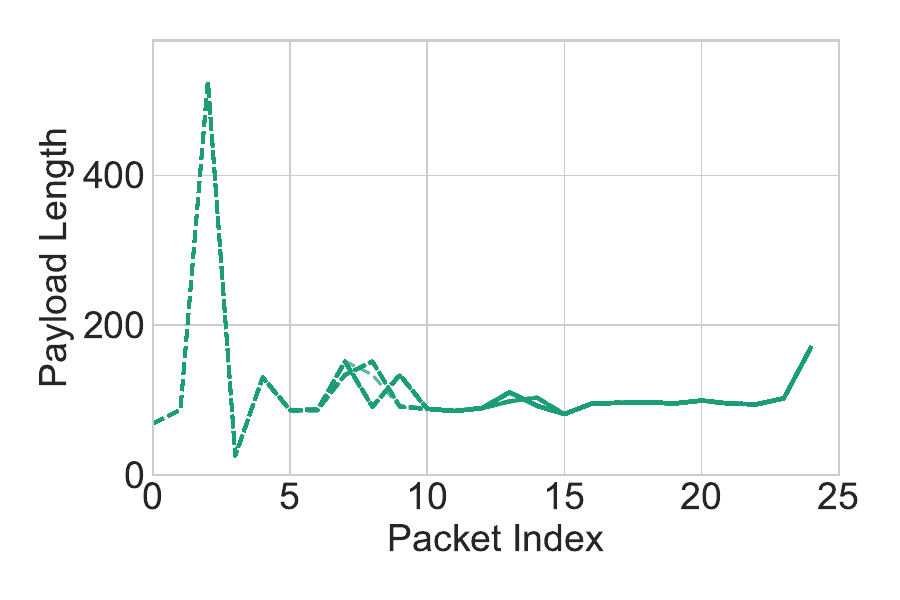}
        \caption{ID: 0.}
        \label{subfig:set-polyas-0}
    \end{subfigure}
    \begin{subfigure}[t]{0.24\linewidth}
        \includegraphics[width=\linewidth]{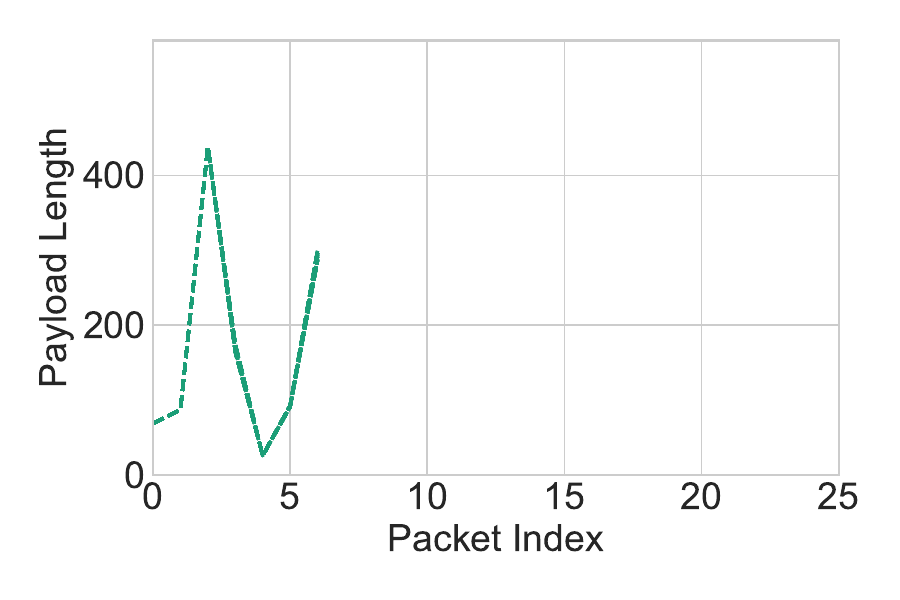}
        \caption{ID: 1.}
        \label{subfig:set-polyas-1}
    \end{subfigure}
    \begin{subfigure}[t]{0.24\linewidth}
        \includegraphics[width=\linewidth]{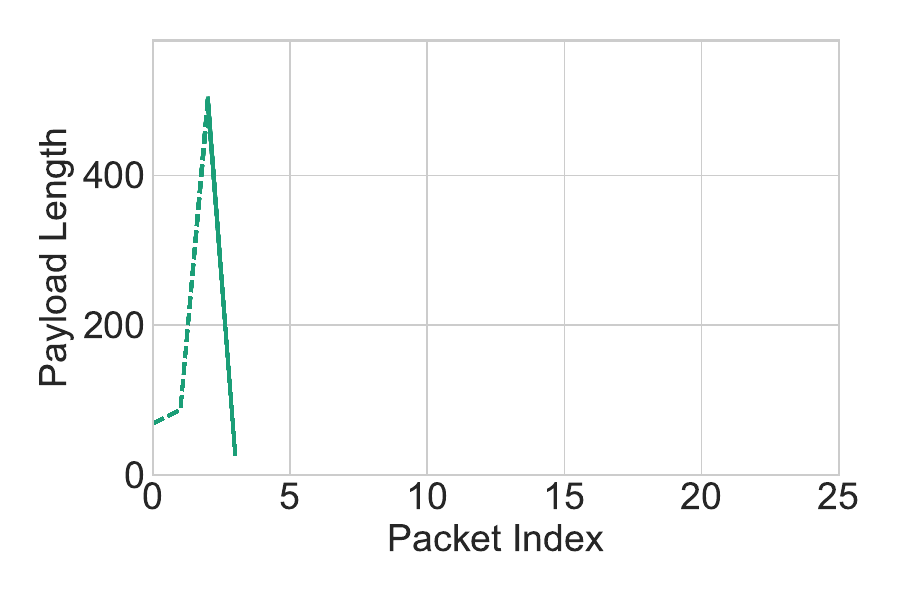}
        \caption{ID: 2.}
        \label{subfig:set-polyas-2}
    \end{subfigure}
    \begin{subfigure}[t]{0.24\linewidth}
        \includegraphics[width=\linewidth]{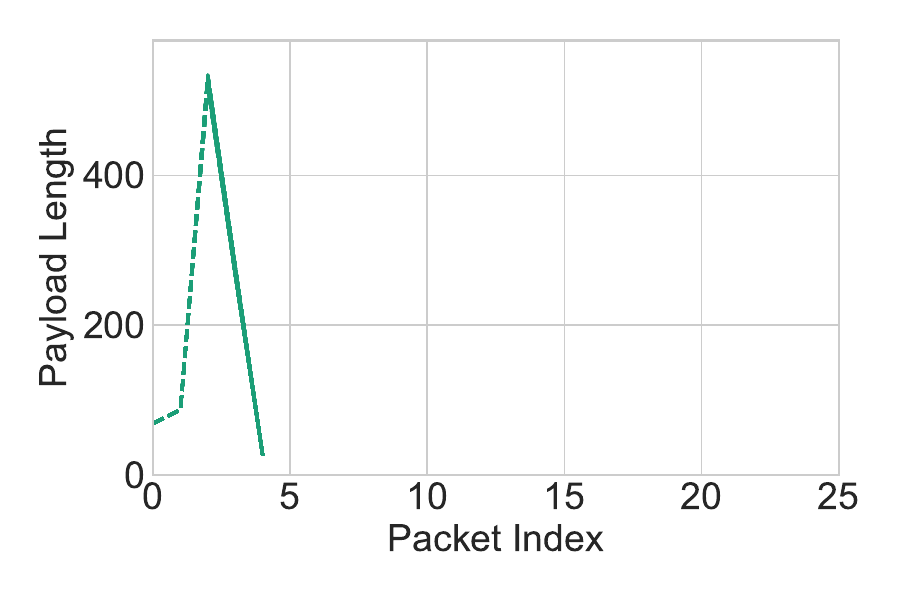}
        \caption{ID: 3.}
        \label{subfig:set-polyas-3}
    \end{subfigure}

    \caption{POLYAS action patterns.}
    \label{fig:patterns-polyas}
\end{figure}

We consider outgoing TLS packets with non-empty application-layer payloads. 
This choice is due to the fact that this class of packets forms patterns of actions that are largely or completely repeated from voter to voter (Fig.~\ref{fig:patterns-eligo} and Fig.~\ref{fig:patterns-polyas}).
The first step is to identify distinct actions within each voter's interaction.
To do this, we apply DBSCAN clustering to each voter's traffic individually (intra-voter mode), using timestamps and inter-arrival times~(IATs) to isolate activity bursts.
DBSCAN is chosen for its ability to discover clusters of arbitrary shape without requiring a predefined number of clusters, and for its robustness to noise, properties that are particularly useful in our setting, where voters may perform actions at irregular intervals.
These features make it more effective then other techniques such as K-Means, which might require predefined cluster counts.

\subsection{Voter Action Classification}
\label{subsec:action}

We develop two models for voter action classification: one based on set theory and the other on clustering techniques.
The set-theoretic method offers near-perfect accuracy by leveraging deterministic rules for identifying packet structures unique to each voter action.
However, this approach is more computationally intensive, particularly when scaling to large datasets or real-time scenarios.
In contrast, the clustering-based model relies on unsupervised learning to group packets with similar structural features.
While its accuracy is slightly lower, it remains sufficiently high for practical use and offers significantly better scalability due to its reduced computational complexity.
Table~\ref{tab:ids} shows the mapping between IDs and actions we aim to identify.
These actions are described as follows.
Note that, for Eligo, a typical logical actions sequence is 0 → 1 → 2 → 4 → 3/7 → 5 → 6, while for POLYAS is 0 → 1 → 2 → 3.\footnote{Eligo's event flow is intentionally reordered so that event ID 3 corresponds to the ``send vote'' action, aligning it with POLYAS for consistency.}
\begin{itemize}
    \item \textbf{Load event:} initial page load or redirection to the voting platform, typically triggered by clicking the election link received by the voter.
    \item \textbf{Log in:} authentication step where the voter enters their credentials to access the voting system.
    \item \textbf{Reload data:} system-side refresh or update of ballot or user session data, often done automatically before vote casting.
    \item \textbf{Open ballot info:} optional step where the voter accesses informational content about the ballot, such as the vote type or maximum number of preferences allowed.
    \item \textbf{Open ballot:} action that loads or displays the actual voting interface, allowing the voter to make selections.
    \item \textbf{Send vote (valid):} submission of a correctly filled ballot, indicating the voter has made a valid choice and cast their vote.
    \item \textbf{Send vote (spoiled):} submission of an intentionally spoiled or blank ballot, indicating voter abstention or protest.
    \item \textbf{Redirect home:} navigation action that redirects the voter to the home or landing page after vote submission.
    \item \textbf{Log out:} final session termination step where the voter exits the platform securely.
\end{itemize}

\begin{table}[!htbp]
\scriptsize
    \centering
    \renewcommand{\arraystretch}{1.2}
    \caption{Action IDs for the voting platforms.}
    \label{tab:ids}
    \begin{subtable}[t]{0.48\linewidth}
        \centering
        \caption{Eligo.}
        \label{subtab:eligo}
        \begin{tabular}{@{}c|l@{}}
            \hline
            \textbf{ID} & \textbf{Action} \\
            \hline
            0 & Load event \\
            1 & Log in \\
            2 & Open ballot info \\
            3 & Send vote (valid) \\
            4 & Open ballot \\
            5 & Redirect home \\
            6 & Log out \\
            7 & Send vote (spoiled) \\
            \hline
        \end{tabular}
        
    \end{subtable}
    \begin{subtable}[t]{0.48\linewidth}
        \centering
        \caption{POLYAS.}
        \label{subtab:polyas}
        \begin{tabular}{@{}c|l@{}}
            \hline
            \textbf{ID} & \textbf{Action} \\
            \hline
            0 & Load event \\
            1 & Log in \\
            2 & Reload data \\
            3 & Send vote \\
            \hline
        \end{tabular}

    \end{subtable}
\end{table}

\paragraph{Set Theory Based Model.}

Once activity bursts are identified, they can be mapped to actions using a set-theoretic model. Its simplicity, transparency, and deterministic nature make it suitable for adversaries without a large-scale training infrastructure.
While less flexible in handling noisy cases, it performs well when distinctive action patterns, such as payload length sequences, are present.
The process is as follows: first, we extract a set of unique actions and their characteristic payload length patterns. Then, previously identified activity bursts are classified based on this mapping.
Let a sequence of $N$ actions (unique or repeated) be performed in an online voting system. Define $\mathcal{S}_i = \{s_{i,j}\}$, where $i \in [0, N-1]$ and $j \in [0, M_i - 1]$, as the set of $M_i$ payload lengths corresponding to the $i$-th action. If two such sets are identical (i.e., $\mathcal{S}_i = \mathcal{S}_k$ for $i \ne k$), we retain only one and discard duplicates. Let $N’ \leq N$ be the number of unique action sets after this reduction.
Next, for each $\mathcal{S}_i$, we compute the subset shared with any other set $\mathcal{T}_i = \mathcal{S}_i \cap \left(\bigcup_{k \ne i} \mathcal{S}_k\right)$.
We also define $\mathcal{T}_i^* = \mathcal{S}_i \cap \mathcal{S}_k^*$, where $k^* = \arg \max_{k \ne i} |\mathcal{S}_i \cap \mathcal{S}_k|$, capturing the maximum overlap with another action.
Based on intersection information, the final set of unique user actions is determined sequentially as follows:

\begin{enumerate}
    \item If $\mathcal{T}_i = \emptyset$, then the corresponding $i$-th~activity burst is a unique action;
    \item If $\left| \mathcal{S}_i\right| = \left| \mathcal{T}_i^*\right|$, then the corresponding $i$-th~activity burst is a unique action;
    \item If $\left| \mathcal{T}_i\right| = \left| \mathcal{T}_i^*\right|$ and there exists $l \neq i, \ {l \in [0, N' - 1]}$ such that  $\mathcal{T}_l^* = \mathcal{T}_i^*$, then activity burst $m^*$ is a unique action where $m^* = \arg \max_{m \in \{i,l\}} {\left|S_m \right|}$ and ${\mathcal{S}_{m^*} = \mathcal{S}_i \cup \mathcal{S}_l}$;
    \item If $\left| \mathcal{T}_i\right| > \left| \mathcal{T}_i^*\right|$, then the corresponding $i$-th~activity burst is a unique action.
\end{enumerate}

\begin{figure*}[!htbp]
    \centering
    \begin{subfigure}[t]{0.24\textwidth}
        \includegraphics[width=\linewidth]{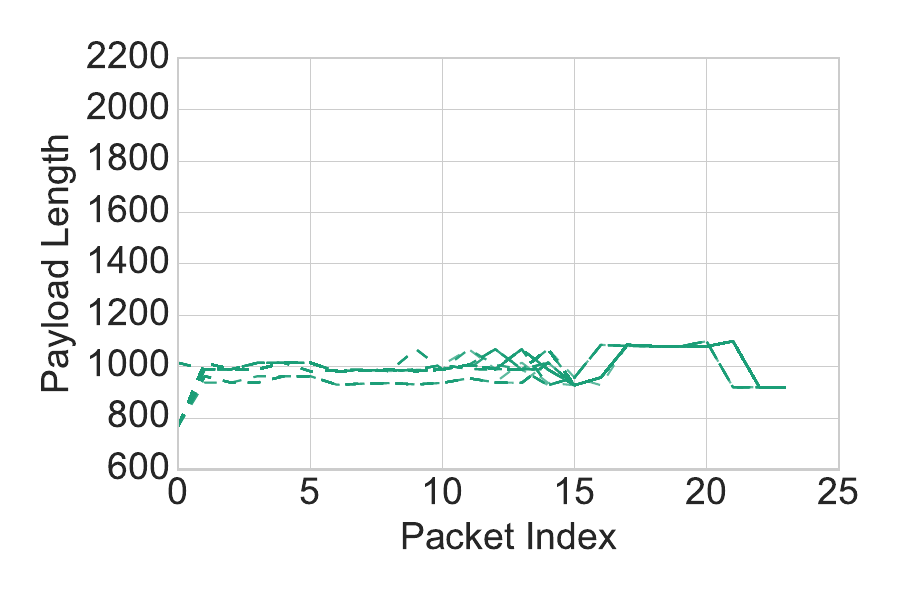}
        \caption{ID: 0.}
        \label{subfig:set-eligo-0}
    \end{subfigure}
    \begin{subfigure}[t]{0.24\textwidth}
        \includegraphics[width=\linewidth]{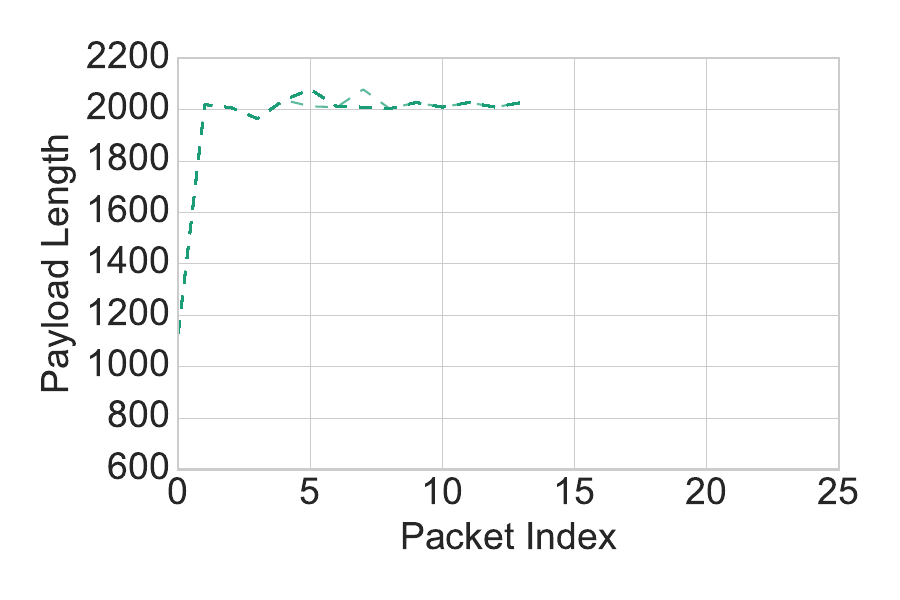}
        \caption{ID: 1.}
        \label{subfig:set-eligo-1}
    \end{subfigure}
    \begin{subfigure}[t]{0.24\textwidth}
        \includegraphics[width=\linewidth]{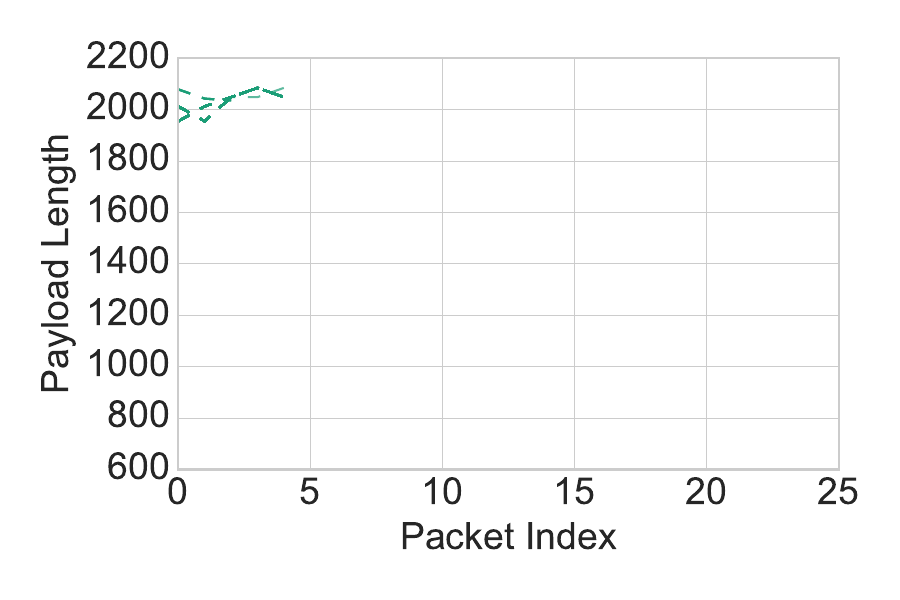}
        \caption{ID: 2.}
        \label{subfig:set-eligo-2}
    \end{subfigure}
    \begin{subfigure}[t]{0.24\textwidth}
        \includegraphics[width=\linewidth]{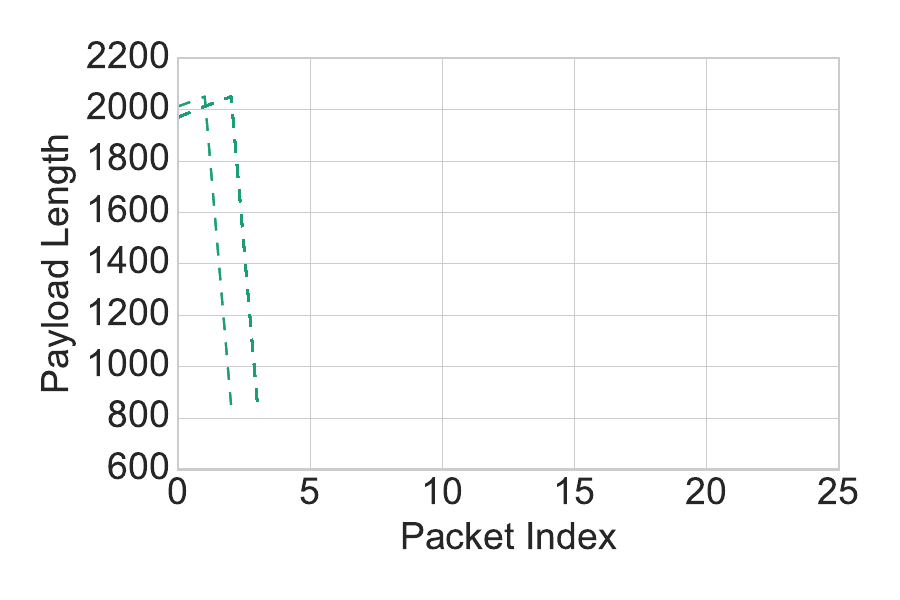}
        \caption{ID: 3.}
        \label{subfig:set-eligo-3}
    \end{subfigure}
    \begin{subfigure}[t]{0.24\textwidth}
        \includegraphics[width=\linewidth]{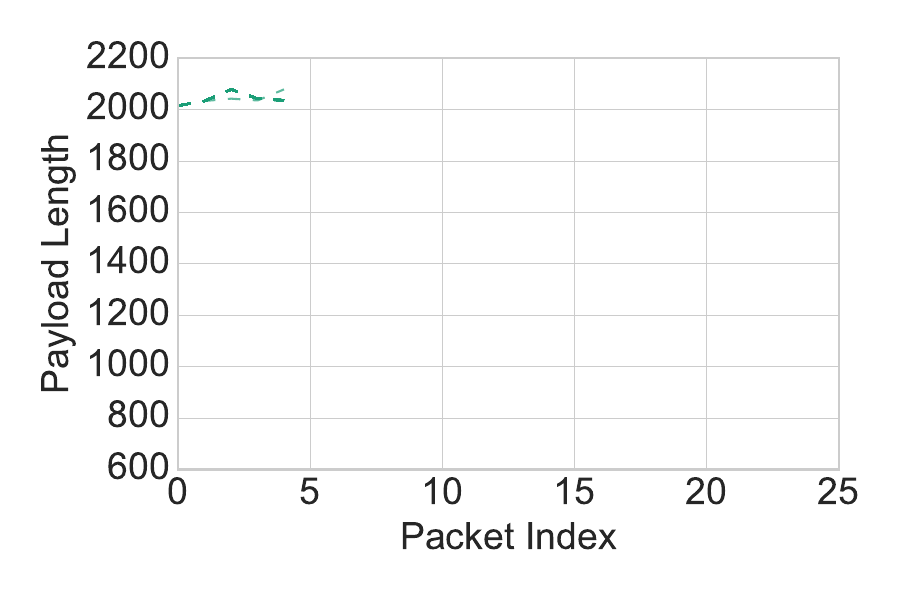}
        \caption{ID: 4.}
        \label{subfig:set-eligo-4}
    \end{subfigure}
    \begin{subfigure}[t]{0.24\textwidth}
        \includegraphics[width=\linewidth]{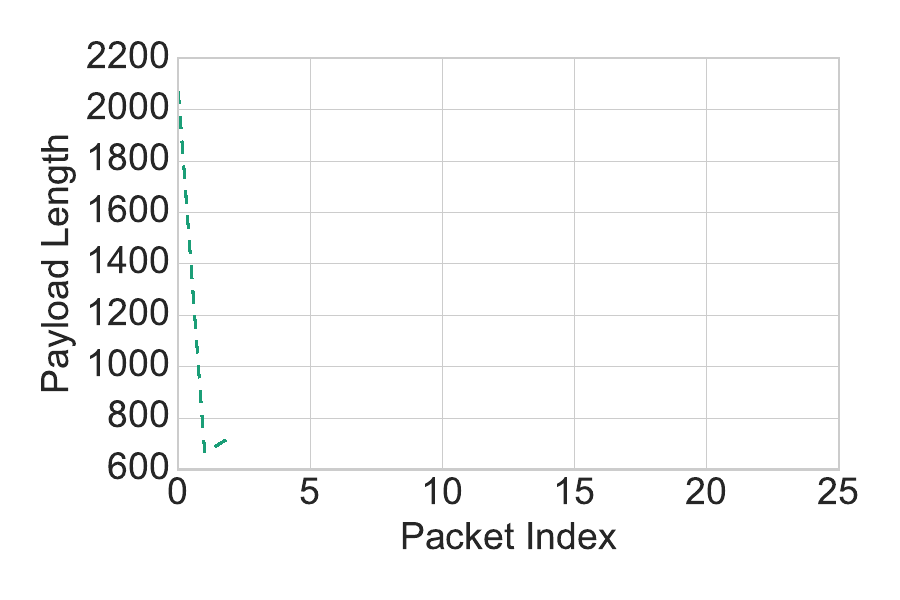}
        \caption{ID: 5.}
        \label{subfig:set-eligo-5}
    \end{subfigure}
    \begin{subfigure}[t]{0.24\textwidth}
        \includegraphics[width=\linewidth]{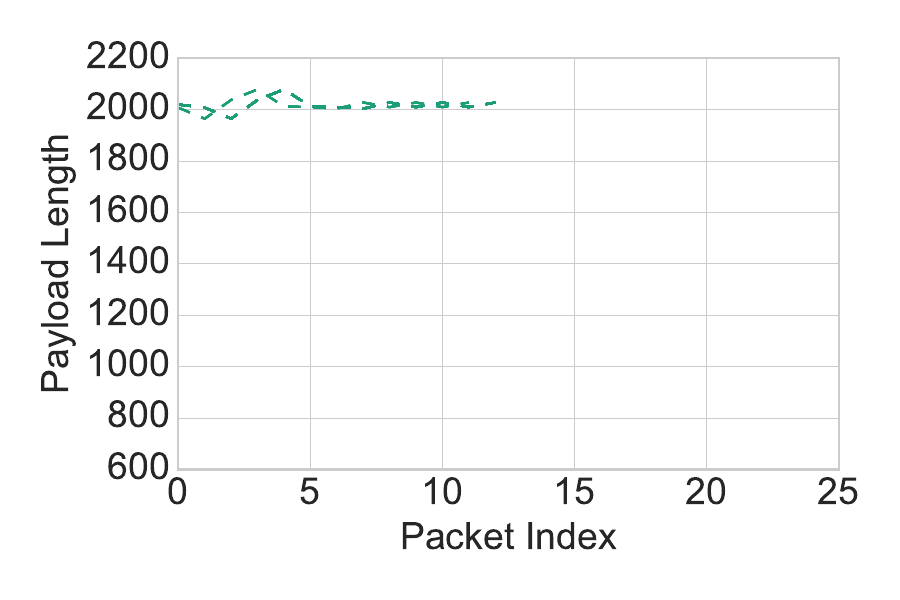}
        \caption{ID: 6.}
        \label{subfig:set-eligo-6}
    \end{subfigure}
    \begin{subfigure}[t]{0.24\textwidth}
        \includegraphics[width=\linewidth]{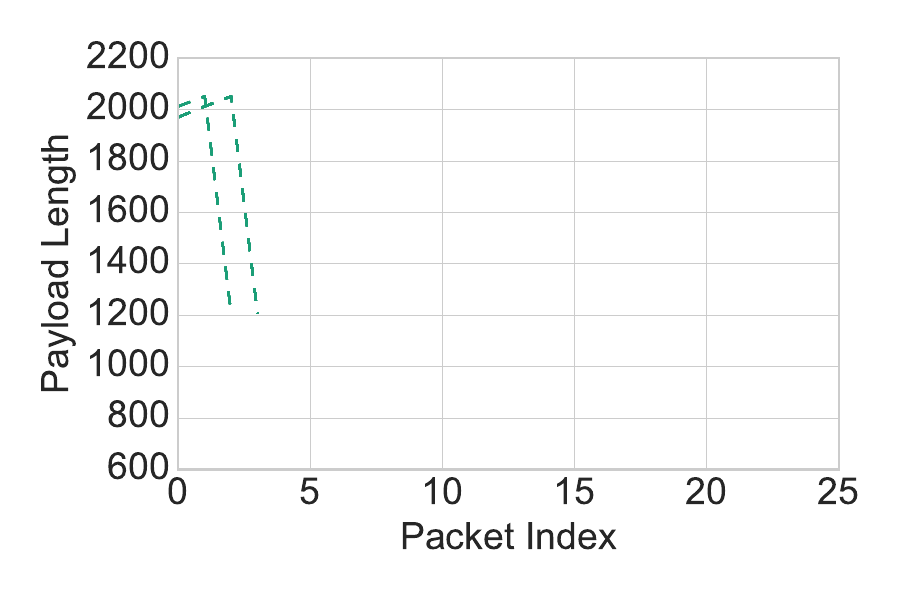}
        \caption{ID: 7}
        \label{subfig:set-eligo-7}
    \end{subfigure}

    \caption{Eligo action patterns.}
    \label{fig:patterns-eligo}
\end{figure*}

Let $N'' \leq N' \leq N$ be the final number of unique user actions.
Suppose a voter performed a sequence of $\hat{N}$ clicks; then $\hat{\mathcal{S}}_p = \{\hat{s}_{p,q}\}$, where $p \in [0, \hat{N}-1]$ and $q \in [0, \hat{M}_p - 1]$, denotes the set of $\hat{M}_p$ payload lengths for the $p$-th activity burst.
This activity burst is classified sequentially as follows:

\begin{enumerate}
    \item If there exists $i \in [0, N''-1]$ such that $\hat{\mathcal{S}}_p = \mathcal{S}_i$, then burst $p$ maps to action $i$.
    \item If $\hat{\mathcal{S}}_p \cap \left(\mathcal{S}i \setminus \cup_{k \neq i}\mathcal{S}_k\right) \neq \emptyset$, then assign burst $p$ to action $i$.
    \item If $\left| \hat{\mathcal{S}}_p \cap \mathcal{S}_i \right| \geq \left|\mathcal{T}_i^* \right| + 1$, then assign burst $p$ to action $i$.
\end{enumerate}

\paragraph{Clustering Based Model.}
Alongside our rule-based method, we implement an unsupervised clustering approach to classify user actions across different voters.
Each activity burst is represented using three features: total payload length, mean payload length, and packet count metrics that capture consistent patterns in TLS-encrypted traffic.
We apply DBSCAN in an inter-voter setting by clustering bursts from all users jointly.
This enables the model to group similar traffic patterns without requiring labeled data.
DBSCAN is chosen for its ability to handle noise and discover clusters of varying shapes without specifying their number in advance.
In Eligo, differences in payload sizes between valid and spoiled votes naturally produce distinct clusters (e.g., action IDs 3 and 7 in Fig.~\ref{fig:patterns-eligo}), allowing us to infer ballot validity directly from the classification step.

\subsection{Vote Submission Detection}
\label{subsec:submission}

As discussed in Section~\ref{system-threat-model}, some attacker roles, such as malicious administrators and passive eavesdroppers, cannot directly observe user interactions or map traffic to specific platform workflows.
Despite these limitations, we demonstrate that it is still feasible to detect vote submission events within encrypted traffic.
Identifying this critical action opens the door to further privacy risks, such as linking voter identities to ballot status.
To achieve this, we propose a method that derives a temporal signature of the vote submission action from aggregated traffic.
This approach exploits consistent structural and timing patterns that emerge across users when the same action is repeated.
The method proceeds as follows:
\begin{enumerate}
    \item \textbf{Input:} Extract all TLS packets belonging to identified activity bursts across users.
    \item \textbf{Grouping:} For each unique action and packet index, collect corresponding packets from all users.
    \item \textbf{Sorting:} Sort each packet group by timestamp normalized per action (starting at 0.0\,s) to capture relative timing within the action.
    \item \textbf{Trend estimation:} Apply a rolling mean to each group to obtain timing trends.
    \item \textbf{Signature generation:} Normalize trend curves and average across packet indices to produce the final action signature.
\end{enumerate}
The action signature captures the temporal distribution of a given action across all voters.
In both systems studied, we observe a consistent distinction: actions before vote submission produce smooth, gradually rising curves, reflecting their spread over time; in contrast, vote submission shows a distinct mid-curve jump, indicating a concentration of events in tighter time intervals.
This distinctive temporal feature, illustrated in Fig.~\ref{fig:action-signatures}, makes vote submissions stand out from other actions.
The jump results from synchronized increases in normalized timestamps across multiple packet indices, which accumulate into a sharp inflection in the averaged action-level curve (Figures~\ref{subfig:sign-eligo-3} and~\ref{subfig:sign-polyas-3}).
In POLYAS, this temporal pattern is visible for both valid and invalid ballots, allowing signature computation on the full dataset.
In Eligo, clustering separates valid and spoiled votes, with the signature jump only present in the valid submission cluster.
Since clustering is based on observable traffic features, a passive attacker can isolate the valid cluster without labeled data and compute the signature accordingly.

\newcommand{\figwidth}{0.24\linewidth}

\begin{figure}[!htbp]
    \centering
    \begin{subfigure}[t]{\figwidth}
        \includegraphics[width=\linewidth]{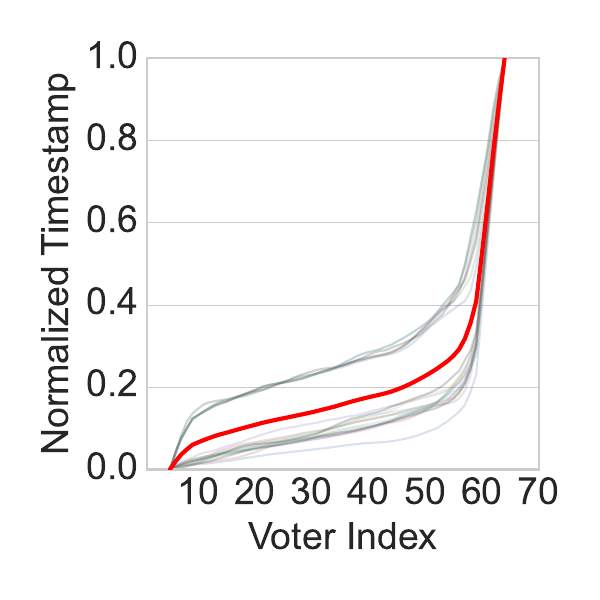}
        \caption{Eligo ID: 0.}
        \label{subfig:sign-eligo-0}
    \end{subfigure}
    \begin{subfigure}[t]{\figwidth}
        \includegraphics[width=\linewidth]{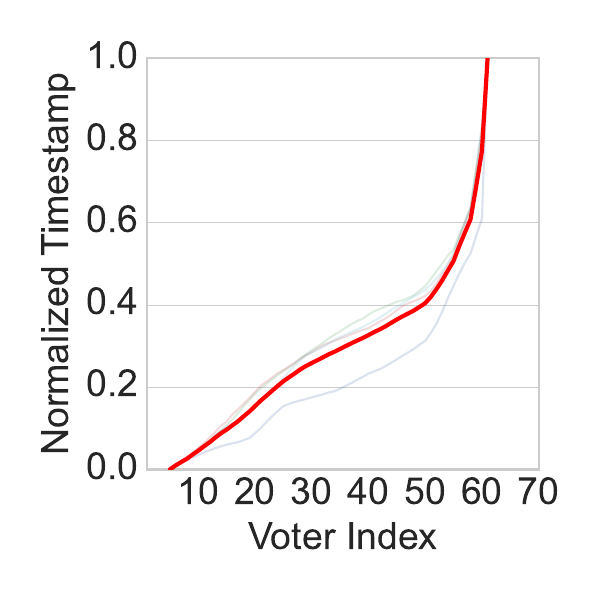}
        \caption{Eligo ID: 1.}
        \label{subfig:sign-eligo-1}
    \end{subfigure}
    \begin{subfigure}[t]{\figwidth}
        \includegraphics[width=\linewidth]{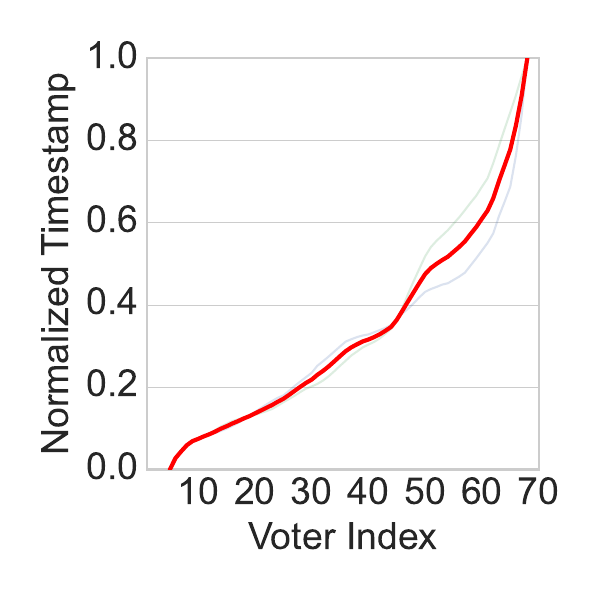}
        \caption{Eligo ID: 2.}
        \label{subfig:sign-eligo-2}
    \end{subfigure}
    \begin{subfigure}[t]{\figwidth}
        \includegraphics[width=\linewidth]{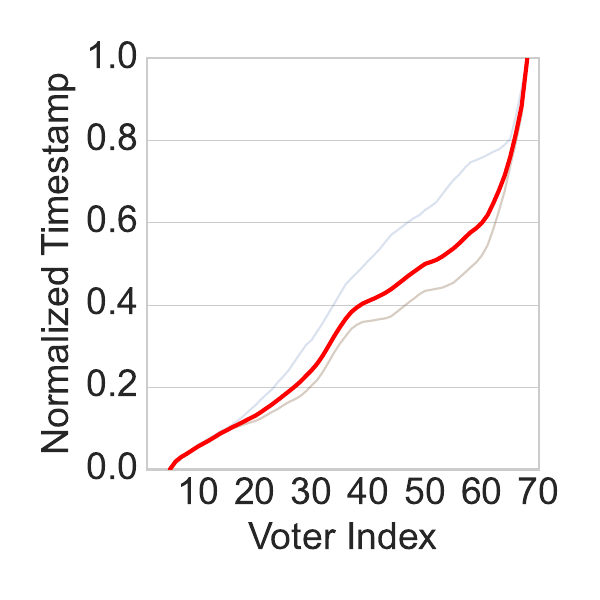}
        \caption{Eligo ID: 3.}
        \label{subfig:sign-eligo-3}
    \end{subfigure}

    \begin{subfigure}[t]{\figwidth}
        \includegraphics[width=\linewidth]{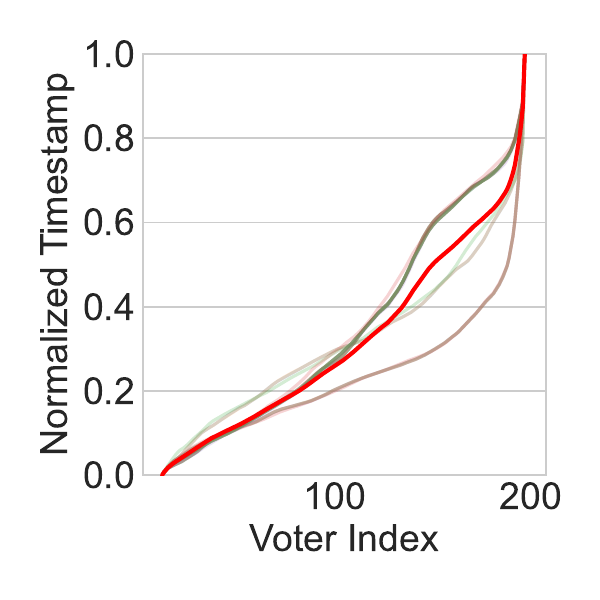}
        \caption{POLYAS ID: 0.}
        \label{subfig:sign-polyas-0}
    \end{subfigure}
    \begin{subfigure}[t]{\figwidth}
        \includegraphics[width=\linewidth]{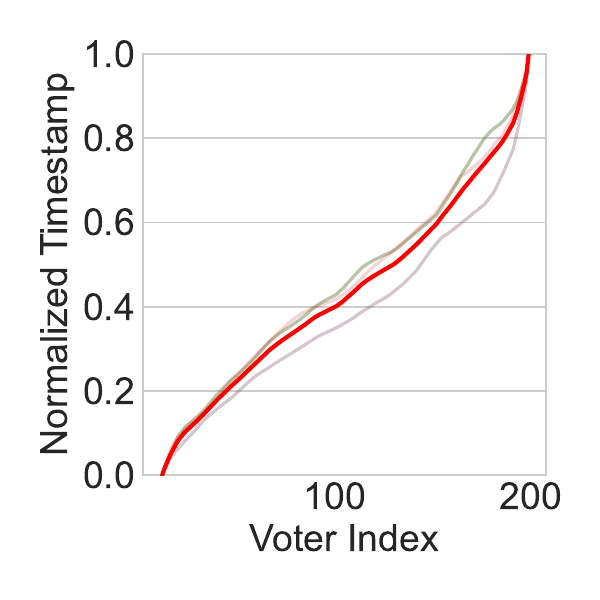}
        \caption{POLYAS ID: 1.}
        \label{subfig:sign-polyas-1}
    \end{subfigure}
    \begin{subfigure}[t]{\figwidth}
        \includegraphics[width=\linewidth]{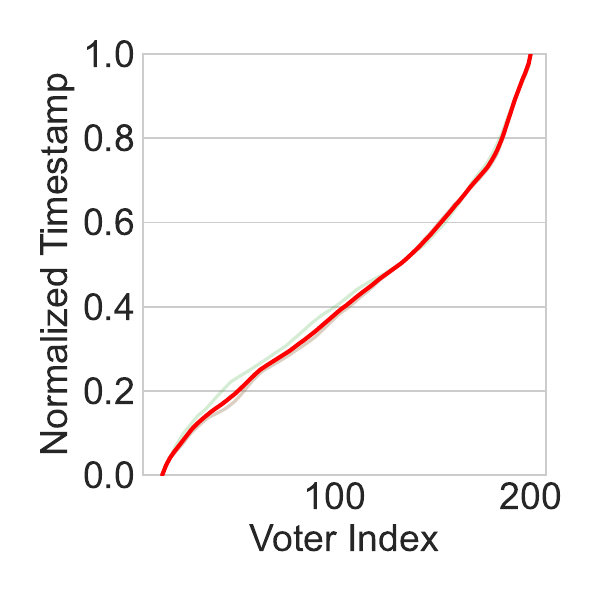}
        \caption{POLYAS ID: 2.}
        \label{subfig:sign-polyas-2}
    \end{subfigure}
    \begin{subfigure}[t]{\figwidth}
        \includegraphics[width=\linewidth]{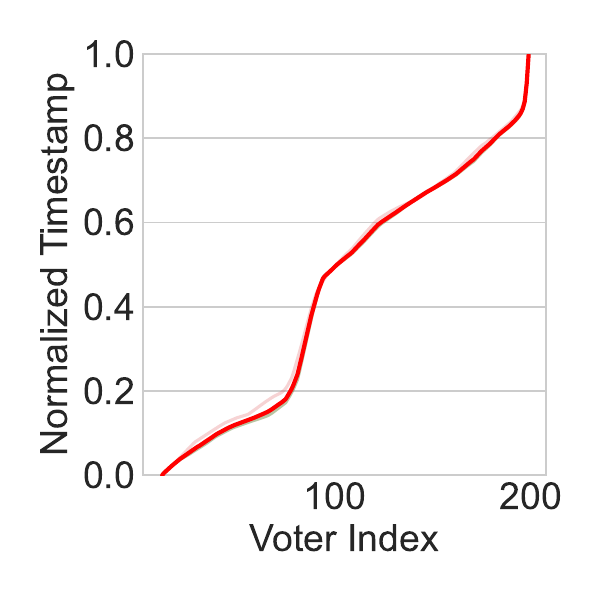}
        \caption{POLYAS ID: 3.}
        \label{subfig:sign-polyas-3}
    \end{subfigure}

    \caption{Action temporal signatures.}
    \label{fig:action-signatures}
\end{figure}

\subsection{Ballot Validity Classification}
\label{subsec:valid}

This stage focuses exclusively on packets corresponding to the ballot submission action, as identified using the methodology described in the previous section.
The goal is to determine whether a submitted ballot is valid or spoiled, based solely on encrypted traffic metadata.

\paragraph{Eligo.}
In Eligo, the distinction between valid and spoiled ballots arises automatically during unsupervised action classification.
This is driven by a consistent, deterministic difference in application-layer payload lengths during ballot submission.
Since these lengths underpin the clustering model, valid and spoiled votes form separate clusters without the need for manual labels.
As a result, ballot validity can be inferred deterministically using straightforward rules based on observed payload patterns.

\paragraph{POLYAS.}
Unlike Eligo, the POLYAS system does not exhibit deterministic differences in traffic metadata between valid and spoiled ballots, making simple rule-based separation infeasible.
To address this, we frame the problem as a binary classification task and apply machine learning techniques to distinguish between the two classes.
We enhance the inter-arrival time (IAT) feature of the ballot submission action using \texttt{tsfresh}~\cite{christ2018time}, a time-series feature extraction toolkit, to capture subtle temporal patterns potentially linked to ballot validity.
Our evaluation spans both tree-based and sequence-aware models suited for detecting nuanced, possibly non-linear differences in structured traffic features. Specifically, we test Random Forest, Gradient Boosting, XGBoost, LightGBM, and a simple ensemble combining Random Forest and LightGBM.
The best ensemble combines a shallow Random Forest (10 estimators, max depth 3) with a lightweight LightGBM (3 estimators, 2 leaf nodes, 0.4 subsampling rate).
Additionally, we train a multi-layer LSTM model to explore whether sequential dependencies in timing are better captured by recurrent architectures.
The LSTM consists of a unidirectional layer (1100 units), a bidirectional layer (380 units), and another unidirectional layer (530 units), followed by dense layers with dropout.
Training is performed for 5 epochs with a batch size of 8 and a learning rate of $10^{-5}$, optimizing for precision.
Model hyperparameters were tuned across multiple runs, selecting configurations that maximized validation precision.
\section{Evaluation}
\label{sec:evaluation}

We now evaluate our methodology and show the results of our attacks by providing details on our experimental setup and numerical results on our attacks' success rates.

\subsection{Experimental Setup}
\label{subsec:setup}

To evaluate our approach, we generate a dataset of 400 simulated voting sessions (200 each on Eligo and POLYAS) with balanced valid and spoiled ballots.
We used balanced datasets to clearly expose learnable distinctions in voting behavior.
While real-world class imbalance may affect classifier performance, standard rebalancing techniques (e.g., class weights, oversampling) can be applied without affecting feasibility.
Indeed, in Eligo, classification relies on fixed payload length patterns, rendering imbalance irrelevant.
User interactions are automated for reproducibility, and network traffic was captured via Wireshark.
It is also worth noting that, although our experiments were conducted in controlled conditions, the primary features we exploit (e.g., inter-arrival times, payload lengths) are artifacts of protocol design rather than user interaction variability.
These features are invariant, making classification robust across different client and network conditions.
After extracting and filtering outgoing TLS packets with non-empty application data using Python, we segment user flows into individual actions using DBSCAN (Section~\ref{subsec:preprocessing}).
Clustering quality is high, with silhouette scores of $0.893\pm0.059$ for Eligo and $0.898\pm0.045$ for POLYAS, and minimal noise, confirming robust action-level segmentation.

\subsection{Voter Action Classification Results}
\label{subsec:action-results}

After segmenting each voter’s traffic into activity bursts using DBSCAN clustering, we applied two approaches to classify the nature of these bursts into specific voter actions: a rule-based model and a clustering-based model.
Both methods operated on the full dataset and did not require labeled data during training. 
Both models are first implemented and evaluated on traffic generated from the Eligo platform and then tested on POLYAS to assess generalizability.
The classification performance is summarized in Table~\ref{tab:acc-AC}.
Both approaches achieve high classification accuracy, with the clustering-based model demonstrating strong generalizability across systems and slightly outperforming the rule-based model on POLYAS.
To further assess the internal consistency and separation quality of the clusters, we evaluated the clustering structure using silhouette scores and noise ratios, which reflect the compactness and distinctiveness of identified clusters. Eligo achieved a silhouette score of~0.958 and a noise ratio of~0.001, indicating highly cohesive and well-separated clusters. Similarly, POLYAS yielded a silhouette score of~0.914 and a noise ratio of~0.001, also demonstrating strong clustering quality.
These results confirm that the clustering model forms highly coherent and separable groupings of user actions across both datasets, reinforcing its applicability in inferring user behavior from encrypted traffic metadata.

\begin{table}[!htbp]
    \scriptsize
    \centering
    \renewcommand{\arraystretch}{1.2}
    \caption{Action classification accuracy for each voting platform. ST = Set Theory model, CL = Clustering model.}
    \label{tab:acc-AC}

    \begin{subtable}[t]{0.48\linewidth}
        \centering
        \caption{Eligo.}
        \label{tab:acc-AC-eligo}
        \begin{tabular}{@{}l|c|c@{}}
            \hline
            \multirow{2}{*}{\textbf{ID}} & \multicolumn{2}{c}{\textbf{Accuracy}} \\ \cline{2-3}
            & \textbf{ST} & \textbf{CL} \\ \hline
            0 & 1.00 & 1.00 \\
            1 & 0.86 & 0.92 \\ 
            2 & 1.00 & 0.99 \\
            4 & 1.00 & 0.82 \\ 
            3/7 & 1.00 & 0.99 \\ 
            5 & 1.00 & 0.94 \\ 
            6 & 1.00 & 0.99 \\ \hline
            \textbf{Avg.} & \textbf{0.98} & \textbf{0.95} \\ \hline
        \end{tabular}
    \end{subtable}
    \begin{subtable}[t]{0.48\linewidth}
        \centering
        \caption{POLYAS.}
        \label{tab:acc-AC-polyas}
        \begin{tabular}{@{}l|c|c@{}}
            \hline
            \multirow{2}{*}{\textbf{ID}} & \multicolumn{2}{c}{\textbf{Accuracy}} \\ \cline{2-3}
            & \textbf{ST} & \textbf{CL} \\ \hline
            0 & 1.00 & 0.99 \\
            1 & 1.00 & 0.98 \\ 
            2 & 0.96 & 0.98 \\ 
            3 & 0.89 & 0.99 \\ \hline
            \textbf{Avg.} & \textbf{0.96} & \textbf{0.99} \\ \hline
        \end{tabular}
    \end{subtable}
\end{table}

\subsection{Ballot Validity Classification}
\label{subsec:validity-results}

The final stage of our analysis focuses on distinguishing between valid and spoiled ballots using encrypted traffic data.
For Eligo, this task is straightforward due to consistent differences in payload lengths during the vote submission action.
These differences lead to clear clustering of valid and spoiled ballots without requiring a separate classifier.
As a result, the system achieves near-perfect classification accuracy of 99.49\% based solely on these deterministic payload patterns.
In contrast, POLYAS does not exhibit such deterministic differences.
To address this, we extract detailed temporal features from packet sequences and apply supervised machine learning models to identify subtle statistical distinctions between ballot types.
We evaluate several classifiers—including Random Forest, Gradient Boosting, XGBoost, LightGBM, an ensemble model (Random Forest + LightGBM), and a multi-layer LSTM—on a dataset with an 80/20 train-test split.
Results (Table~\ref{tab:acc-BC-polyas}) show that LightGBM and the Ensemble model achieve the highest overall performance, with accuracies up to 78\%, along with balanced precision and recall.
The LSTM model demonstrates the highest recall, effectively detecting spoiled ballots, but at the cost of reduced precision.
Gradient Boosting, while showing perfect precision, suffers from very low recall, making it unsuitable when detecting all spoiled ballots is critical.
Overall, while Eligo’s deterministic patterns enable trivial ballot validity classification, our results show that even encrypted traffic from non-deterministic systems like POLYAS can leak enough temporal information to enable classification, revealing a learnable side channel exploitable with up to 78\% accuracy.

\begin{table}[!htbp]
    \scriptsize
    \centering
    \renewcommand{\arraystretch}{1.2}
    \caption{Spoiled ballot identification performance.}
    \label{tab:acc-BC-polyas}
    \begin{tabular}{l|c|c|c}\hline
        \textbf{Classifier} & \textbf{Prec.} & \textbf{Rec.} & \textbf{Acc.} \\ \hline
        Random Forest & 0.56 & 0.90 & 0.60 \\ 
        Gradient Boosting & \textbf{1.00} & 0.20 & 0.60 \\
        XGBoost & 0.80 & 0.40 & 0.65 \\
        LightGBM & 0.73 & 0.80 & 0.75 \\
        Ensemble (RF + LGBM) & 0.74 & 0.85 & \textbf{0.78} \\
        LSTM & 0.68 & \textbf{0.95} & 0.69 \\
        \hline
    \end{tabular}
\end{table}
\section{Discussion}
\label{sec:discussion}

Section~\ref{sec:evaluation} presented our attacks' accuracy and success rates; here, we analyze their implications for each system: Section~\ref{subsec:disc1} covers key factors in identifying vote submissions in Eligo, and Section~\ref{subsec:disc2} offers a statistical analysis of POLYAS under attack.

\subsection{Eligo Vote Submission Identification}
\label{subsec:disc1}

Applying our vote submission detection method to the Eligo dataset reveals a clear difference in temporal signature shapes between valid and spoiled ballots.
Valid submissions show a distinct midsection jump, reflecting tightly coupled backend processes like ballot validation and cryptographic sealing that occur consistently across sessions, causing aligned bursts in traffic.
In contrast, spoiled ballots follow a more variable timing pattern due to early termination or altered processing paths, resulting in smoother signatures without sharp inflections.
This demonstrates how deterministic system behaviors can be inferred passively from encrypted traffic, exposing privacy risks despite the lack of visible payload or labels.

\subsection{POLYAS Ballot Validity Classification}
\label{subsec:disc2}

\begin{figure}[!htbp]
    \centering
    \begin{subfigure}[t]{.49\linewidth}
        \includegraphics[width=\linewidth]{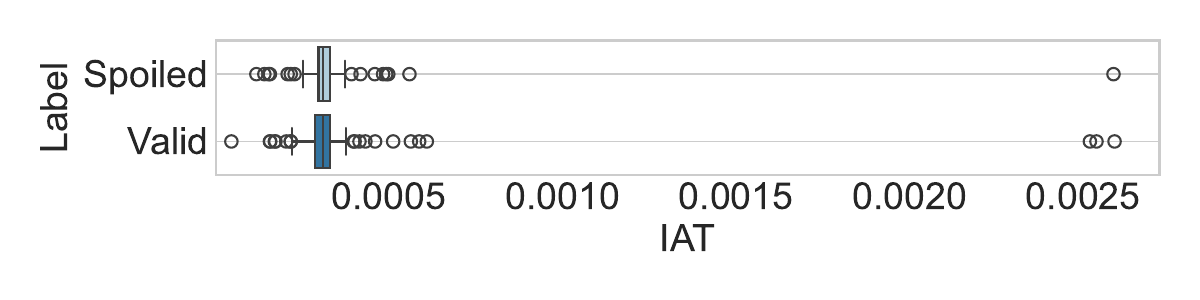}
        \caption{ID: 1.}
        \label{subfig:boxplot-polyas-1}
    \end{subfigure}
    \begin{subfigure}[t]{.49\linewidth}
        \includegraphics[width=\linewidth]{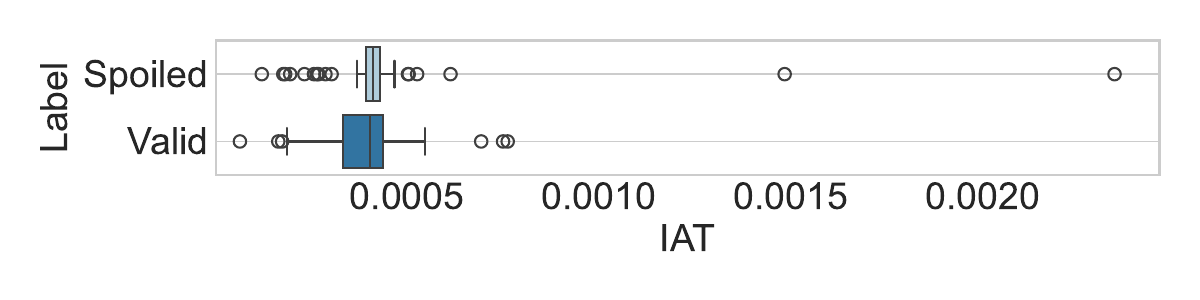}
        \caption{ID: 2.}
        \label{subfig:boxplot-polyas-2}
    \end{subfigure}
    \begin{subfigure}[t]{.49\linewidth}
        \includegraphics[width=\linewidth]{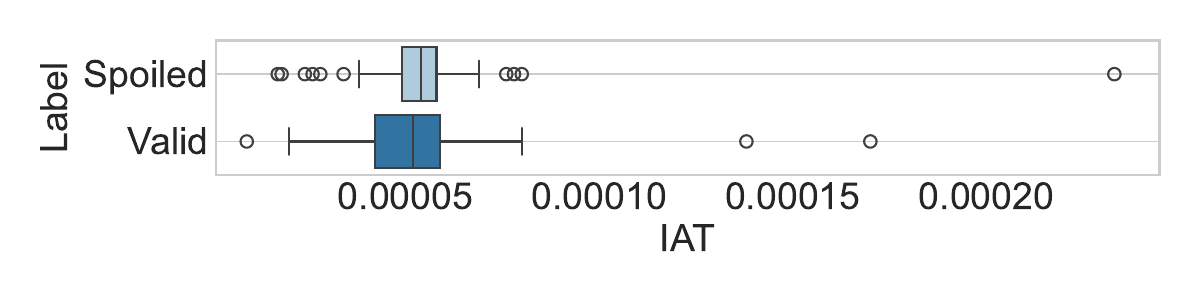}
        \caption{ID: 3.}
        \label{subfig:boxplot-polyas-3}
    \end{subfigure}
    \begin{subfigure}[t]{.49\linewidth}
        \includegraphics[width=\linewidth]{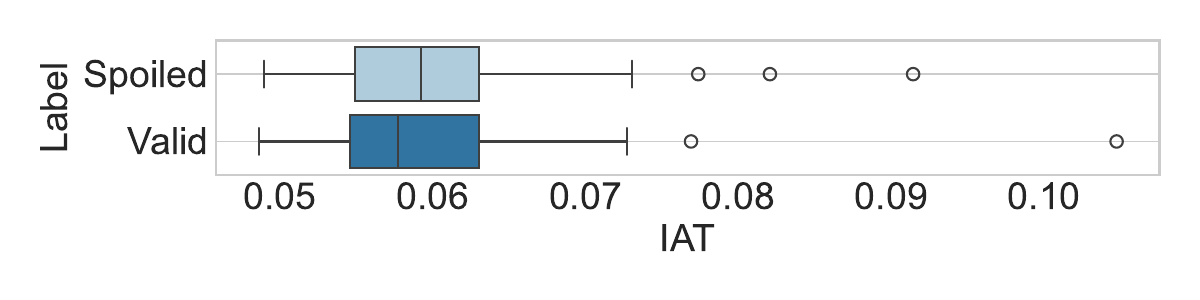}
        \caption{ID: 4.}
        \label{subfig:boxplot-polyas-4}
    \end{subfigure}

    \caption{Comparison of IAT distributions for ballot submission action by packet index.}
    \label{fig:boxplots-polyas}
\end{figure}

Experimental results show that machine learning and deep learning models perform worse on POLYAS than on Eligo, where vote validity was classified with near-perfect accuracy.
This prompts the question of whether POLYAS lacks class-separability or if subtle timing differences go undetected by current models. To investigate, we perform a statistical analysis of IAT distributions in outgoing TLS traffic.
We apply various tests grouped by their focus: \textit{location tests} assess central tendency ($p < 0.05$ suggests linear separability), \textit{scale tests} like Ansari-Bradley evaluate dispersion without assuming normality, \textit{distribution equality} tests such as Kolmogorov-Smirnov detect subtle distribution shifts~\cite{stat-test-ex-KS}, and \textit{location and scale} tests like Mann-Whitney and Lepage capture changes in both mean and variance~\cite{stat-test-ex-MW-Lpg}.
Fig.~\ref{fig:boxplots-polyas} summarizes these findings.
Because each action emits a fixed sequence of TLS packets with payloads, we align IATs by packet index to enable detailed, packet-level comparisons between valid and spoiled ballots, exposing action-specific temporal patterns.

Consistent significance was observed only for the vote submission action, particularly across multiple packet indices and non-parametric tests with $p$-values below 0.05, which capture complex distributional differences. 
Boxplots in Fig.~\ref{fig:boxplots-polyas} further highlight distinct IAT patterns between the two classes at corresponding packet indices.

\begin{table}[!htbp]
    \scriptsize
    \centering
    \renewcommand{\arraystretch}{1.2}
    \caption{P-values of statistical tests for ballot submission action by packet index.}

    \begin{tabular}{l|l|cccc} \hline
    \multirow{2}{*}{\textbf{Objective}} & \multirow{2}{*}{\textbf{Test}} & \multicolumn{4}{c}{\textbf{P-Value}} \\ \cline{3-6}
    & & \multicolumn{1}{c}{\textbf{ID 1}} & \multicolumn{1}{c}{\textbf{ID 2}} & \multicolumn{1}{c}{\textbf{ID 3}} & \multicolumn{1}{c}{\textbf{ID 4}} \\ \hline
    \multirow{1}{*}{Scale}
    & Ansari-Bradley \cite{stat-test-AB} & 0.1266 & \textbf{0.0006} & \textbf{0.0015} & 0.7937 \\ \hline

    \multirow{3}{*}{\begin{tabular}[l]{@{}l@{}}Distribution\\equality\end{tabular}}
    & Cramer--von Mises \cite{stat-test-CvM} & 0.3942 & \textbf{0.0318} & \textbf{0.0358} & 0.7280 \\
    & Epps-Singleton \cite{stat-test-ES} & 0.3223 & \textbf{0.0041} & \textbf{0.0023} & 0.6949 \\
    & Kolmogorov-Smirnov \cite{stat-test-KS} & 0.1994 & \textbf{0.0066} & 0.0524 & 0.7510 \\ \hline

    \multirow{3}{*}{\begin{tabular}[l]{@{}l@{}}Location\\\& Scale\end{tabular}}
    & Cucconi \cite{stat-test-Cucconi} & 0.1678 & \textbf{0.0039} & \textbf{0.0039} & 0.7832 \\
    & Lepage \cite{stat-test-Lpg} & 0.2857 & \textbf{0.0499} & \textbf{0.0430} & 0.2537 \\
    & Podgor-Gastwirth \cite{stat-test-PG} & 0.1599 & \textbf{0.0030} & \textbf{0.0025} & 0.7880 \\ \hline

    \end{tabular}
    \label{tab:tests-res}
\end{table}

In contrast, location-based tests did not yield consistent significance, suggesting that class differences lie not in average behavior but in distributional patterns of packet IATs during vote submission.
While classification accuracy for POLYAS was modest, statistical analysis still reveals class-dependent timing differences, indicating that more tailored models could achieve higher accuracy and pose greater risks to vote secrecy.
\section{Countermeasures}
\label{sec:countermeasure}

We evaluate two countermeasures against traffic analysis, addressing payload length and temporal signatures, through a case study on Eligo, where attacks were most effective and traffic allowed controlled simulation.
For POLYAS, only partial mitigation analysis was possible due to its architectural and traffic differences.

\paragraph{Payload Padding.}

To defend against payload-based inference, we simulated padding all non-empty application data packets to a uniform maximum length of 2085 bytes, the largest observed payload.
This increased RAM usage by up to 109\% in the worst case and 32\% on average (Table~\ref{tab:cm-eligo}).
After padding, ballot validity classification accuracy dropped to 0\%, as deterministic packet length differences between valid and spoiled ballots were eliminated, showing that simple payload padding effectively counters this attack.
While POLYAS action classification also depends on payload lengths and would be blurred by padding, its ballot validity classification relies on temporal features and remains unaffected by this mitigation.

\paragraph{Time Equalization.}

To counter timing-based inference, we propose equalizing IATs of application data packets within each action to flatten temporal signatures and block timing-based attacks.
For Eligo, we simulate this by sampling IATs from a baseline distribution of maximum-length packets and applying them to a synthetic voter, increasing action durations by up to 6.11 seconds (2.06 seconds on average).
This uniform timing removes the distinctive signature jump of vote submissions, effectively preventing detection.
For POLYAS, the lack of a clear link between payload size and IATs makes realistic time equalization simulation infeasible, so we do not report similar results, though system-specific timing equalization remains a promising approach.

\begin{table}[!h]
    \scriptsize
    \centering
    \renewcommand{\arraystretch}{1.2}
    \caption{Mitigation evaluation for Eligo}
    \begin{tabular}{l|c|c}\hline
        \textbf{Action} & \textbf{\makecell{Payload \\ Increase}} & \textbf{\makecell{Action Duration \\ Increase (seconds)}} \\ \hline
        
        Load voting event page & 1.09 & 5.49 (6.11) \\
        Log in & 0.07 & 3.03 (3.85) \\ 
        Open ballot info & 0.02 & 2.36 (0.98) \\
        Open ballot content & 0.03 & 2.37 (1.17) \\ 
        Send vote & 0.21 & 5.14 (1.06) \\ 
        Redirect to home page & 0.02 & 2.35 (0.92) \\ 
        Log out & 0.81 & 1.49 (0.30) \\ \cline{1-3}
        \textbf{Avg.} & \textbf{0.32} & \textbf{3.18 (2.06)} \\ \hline

    \end{tabular}
    \label{tab:cm-eligo}
\end{table}
\section{Conclusions}
\label{sec:conclusions}

This paper shows that key voting actions, such as ballot submission, can be inferred from encrypted TLS traffic using timing and payload features, without needing system access.
Our methodology detected vote submissions and classified ballot validity in two real-world systems.
We also proposed mitigations, payload padding, and time equalization, which significantly reduce inference success, underscoring the need to treat traffic metadata as a serious privacy concern in online voting systems.

\paragraph{Limitations.}
Our methodology, developed using Eligo, was also applied to POLYAS without system-specific tuning to assess its generalizability.
While POLYAS yielded lower classification accuracy, statistical analysis revealed class-dependent timing patterns, indicating potential for improvement through tailored models.
Future work could explore system-specific strategies to boost performance further.

\paragraph{Future Works.}
Looking forward, our findings suggest several future directions.
Our current work assumes passive adversaries who do not alter traffic, but future research should explore active threat models where attackers manipulate or inject data to amplify leakage or disrupt voting, along with defenses against such attacks.
Additionally, these methods could be extended to infer more detailed privacy risks, like voters’ preferences, emphasizing the urgent need for robust traffic-level privacy protections in online voting systems.

\begin{credits}
\subsubsection{\ackname} We thank Eligo for granting us access to their voting system, which enabled the experimental evaluation presented in this work.
\end{credits}

\bibliographystyle{splncs04}
\bibliography{bibliography}

\end{document}